\documentclass[12pt]{article}
\usepackage[numbers]{natbib}
\usepackage{hyperref}

\usepackage{amstext}
\usepackage{color}
\usepackage{float}
\restylefloat{table}

\renewcommand\L{LRR}
\renewcommand\S{SRR}
\newcommand\hs{\hskip 15pt}

\title{Long range versus short range aerial transmission of SARS-CoV-2
}

\author{John E. M\raise.5ex\hbox{c}Carthy
\thanks{Partially supported by  National Institutes of Health Grant R01 AG052550-01A1 and
 National Science Foundation Grant
DMS 1565243}
\\Washington University in St. Louis
\and
Myles T. M\raise.5ex\hbox{c}Carthy
\\ University of Illinois at Urbana-Champaign
\and
Bob A. Dumas
\\Omnium LLC
}

\date{\today}

\begin{document}

\maketitle

Corresponding author:
John McCarthy,  Department of Mathematics and Statistics,
Washington University in St. Louis,
1 Brookings Drive, St. Louis MO 63130

{\em email:} mccarthy@wustl.edu

\begin{abstract}
We compare the quantitative risk of infection from short range airborne transmission of 
SARS-CoV-2  to the long-range risk from aerosol transmission in an enclosed space.
We use mathematical techniques to show that the risk is approximately inversley proportional to the ventilation rate per person in the space. To estimate the constant of proportionality, we use recent work by Chen et al.
\cite{chen20} modelling the viral load received at a distance of 1 meter and compare this to estimates of the 
total aerosolized viral load.
\end{abstract}


\bibliographystyle{plainnat}

\section{Introduction}

There are believed to be four transmission routes for severe acute respiratory syndrome-coronavirus 2 (SARS-CoV-2): touching an infected person, fomites (touching an infected surface), aerial transmission by droplets, and
aerial transmission by aerosols. 
Recently there have been growing calls to take aerosol transmission into account when assessing 
COVID-19 risks indoors \cite{mor20}, and on 9 July 2020, the W.H.O.  said \cite{WHOaerosol}
``There have been reported outbreaks of COVID-19 in some closed settings, such as restaurants, nightclubs, places of worship or places of work where people may be shouting, talking, or singing.  In these outbreaks, aerosol transmission, particularly in these indoor locations where there are crowded and inadequately ventilated spaces where infected persons spend long periods of time with others, cannot be ruled out.  More studies are urgently needed to investigate such instances and assess their significance for transmission of COVID-19.'' 

It is hard to quantitatively compare risks from different routes.
There have been a small number of studies,  such as \cite{Buon20}, that fit parameters to some
specific events where a large number of infections are known to have occurred. However, it is hard to 
account for the sample bias here---if one analyzes only super-spreader events, estimates of infection risk
will be inflated.

The purpose of this note is to try to compare the relative viral loads that somebody would receive
from exposure to an infectious person standing close to them and the amount they would receive
 via aerosols in an indoor environment where they remain for a long time (such as a class-room, airplane, 
 indoor stadium).
 These can be thought of as short-range and long-range aerial transmission.
 For convenience, we shall refer to short-range transmission as droplet borne, and long-range as aerosol borne,
 even though aerosols contribute to the short range transmission as well \cite{chen20}.
 Our main conclusion (see equation (\ref{eqlr}) in Section \ref{secaer}) is that the long-range risk is approximately
 proportional to $1/F$, where $F$ is the total volume of air exchanged by the ventilation system per person per minute (and it is essential that the air removed is either vented or disinfected, not merely recirculated with the virus still in it).
 Indeed, from equation (\ref{eqlr}) we have that $\L$, the long-range risk, in terms of virions inhaled per minute, is approximately
 \begin{equation}
 \label{eqlrv}
 \L \ \approx \ 
\frac{Iv }{F} \pi,
 \end{equation}
 where $\pi$ is the proportion of the people in the enclosure who are infectious, $I$ is the breathing rate per minute, and $v$ is the number of virions per minute that an infectious person exhales that become aerosolized.

\vskip 10 pt
Currently, $v$ is not known, nor is the risk of infection arising from inhaling a given viral load.
So to compare (\ref{eqlrv}) to the short range risk $\S$, the risk from direct exposure to a person 1m away, even approximately, we need 
to make  some assumptions. We shall use the results of \cite{chen20}, where the authors estimate what
fraction of exhaled saliva is inhaled by another person standing one meter away, face-to-face. 
We shall then estimate what fraction of exhaled saliva becomes aerosolized, and look at the ratio.
 Our conclusion (equation (\ref{eq2})) is that the relative risk $\rho$
 of long-range aerosol transmission to short-range droplet exposure from one person 1 meter away
 is roughly $c/F$, where $F$ is measured in liters per person per minute, 
 \begin{equation}
\label{eqma}
\rho \ = \ \L/\S \ \approx \ \frac{c}{F} .
\end{equation}
We estimate $c$ to be between 6 and 2000, provided that infected people are not coughing or sneezing (see Section \ref{secquant}).
 The numerator increases dramatically if infectious people are coughing or sneezing. Sneezing once per minute increases the aerosol load by a factor of around 50, in addition to increasing the short-range risk.
 
We emphasize that while we think the functional form of  (\ref{eqma}) is approximately correct --- the risk will vary like the reciprocal of the venilation rate per person --- we are not confident that the numerical value of $c$ that we estimate
is accurate, even to an order of magnitude. It will take experimentation to find a good value for the numerator.

\section{Aerosols}
\label{secaer}

Let $v$ be the number of aerosolized virions
 produced by one infectious  person per minute.
Let $A(t) $ be the total number of virions in aerosol form in the air at time $t$.
Let $k$ be the fraction of air removed every minute.
 With complete mixing, we
would get
\[
\frac{dA}{dt} \ = \ v - kA ,
\]
whose solution is
\[
A(t) = \frac{v}{k} ( 1 - e^{-kt}).
\]
So we shall assume that an 
upper bound for $A(t)$ is the equilibrium value of $\frac{v}{k}$.

Let $V$ be the  volume of the indoor region (in liters).
The concentration dispersed in the air is then bounded by
\[
\frac{v}{kV} .
\]
Note that $kV$ is the total volume of air extracted per minute; let us call this $E$.
Let $I$ be the volume of air a person inhales per minute (approximately 10 liters).
We conclude that an approximate upper bound on how much a 
person inhales from a long-range source is
\[
\frac{Iv}{E}
\]
Let $n_t$ be the number of people in attendance, and $n_i$ the number of those who are infected.
Let $\pi = n_i/n_t$ be the fraction of attendees who are infectious.

We get that the long-range risk is then
\[
\L \ = \ 
\frac{I v}{E} n_i
\]
Let $F = E/n_t$ be the volume of air exchanged per person per minute.
We conclude that
\begin{equation}
\label{eqlr}
\L \ = \ 
\frac{Iv }{F} \pi .
\end{equation}
 
 A more sophistiated analysis of aerosol infection risks has been done in 
 \cite{Buon20}.

\section{Droplets}
\label{secdrop}
When an infected person exhales (or coughs or sneezes) they spray droplets of saliva that contain the virus into the air. The larger ones remain as droplets, and either collide with a person, land on a surface, or fall to the ground. The smaller ones become aerosolized and waft through the air, where they can remain for hours.
During this time they can travel a long distance, and potentially be inhaled by a distant person.

We shall assume that the viral concentration in droplets is independent of their size.
Small droplets will rapidly evaporate, but since we wish to compare the viral load of aerosolized droplets to
that of  larger droplets, when we speak of volume we will always mean the volume of the droplet at the time of exhalation.

In \cite{chen20}, the authors analyze short-range droplet transmission, based on a mathematical model.
They argue that there are two main routes of short-range non-fomite transmission: large droplets that are projected directly into the mouth, nose and eyes of a nearby facing person at the same height (they ignore droplets that hit any other part of the face or body), and small droplets that
enter the air stream and are inhaled. Large droplets are intrinsically short-range, because gravity pulls
them down to the floor in a short period of time. (If coughed out, they can travel a long distance horizontally however).
They conclude that mid-size droplets (defined as having initial diameters 75-400 $\mu$m)
fall to the ground within a meter. Smaller droplets follow the air-stream; larger ones decelerate horizontally more slowly, so travel farther, but will settle to the ground.
Moreover, they conclude that at distances over 0.3m (talking) and over 0.8m (coughing) the majority
of exposure comes from inhaled droplets rather than deposited droplets.

For their base data, \cite{chen20} use a paper by Duguid \cite{dug46} who measured the number and size of droplets
exhaled by a person coughing, and by counting loudly from 1 to 100. The latter produced a total
measured  volume of saliva of 0.36 $\mu$L (of which $2 \times 10^{-3}$ $\mu$L came from droplets with a diameter less than $75 \mu$m).  The conclusion of \cite{chen20} that we 
will use as an anchor for short-range aerial transmission is that 
{\em face-to-face, at a range of 1m, a person inhales $6.2 \times 10^{-6}$ $\mu$L 
of the original 0.36 $\mu$L of the talking emission, almost all of it from droplets smaller than 75 $\mu$m.  }

To calculate the volume that becomes aerosolized, we shall  use Duguid's measurements of what fraction
of droplets of a give size remain airborne after 10 minutes; this was 88\% for droplets 0.5-1 $\mu$m, 
7\% for 1-2 $\mu$m, and down to 11\% for droplets 8-10 $\mu$m in diameter.

%

\section{Long range risk comparison to short range}
\label{secquant}
How do we compare $\L$ from (\ref{eqlr}) to the short-range risk from Section \ref{secdrop}?
Instead of $v$, let us work with $b$, the volume of  droplets a person exhales per minute that become
aerosolized. Then we want to compare
\[
\frac{Ib }{F} \pi
\]
to the volume of droplets inhaled from direct transmission from a meter away.

Let us make some numerical assumptions.
\begin{enumerate}
\item
Duguid doesn't record the time it takes to count loudly to 100; we shall assume   it took 100 seconds = 1.67 minutes.
So we shall assume that the amount of saliva inhaled by the direct route at 1 meter is $0.6$ times
the figure from \cite{chen20}, i.e. $3.7 \times 10^{-6}$  $\mu$L/min.
\item
Duguid measures the number and size of droplets emitted by a sneeze, and how many of them remain after 10, 20, 30, and more minutes. We shall assume that the proportion of droplets that remain after 10 minutes are the ones that become aerosolized. This is of course an over-estimate, since some of these will settle out in the next 10 minutes.
While 88\% of droplets of diameter 0.5-1 $\mu$m remained airborne after 10 minutes, only 11\% of
those with diameters 8-10 $\mu$m remained, and none larger. Applying these percentages to Duguid's size distributions of droplets produced by counting, we would get that $3.5 \times 10^{-6}$ $\mu$L of the 
$0.36 $ $\mu$L of total volume would become aerosolized, or $2.1 \times 10^{-6}$ $\mu$L/minute.
We will therefore use this  estimate for $b$:
\[
b \ = \ 2.1 \times 10^{-6} \mu{\rm L/min}.
\]
\end{enumerate}

With these assumptions, and taking $I = 10$ L/min, we get
\[
\L \ = \ \frac{2.1 \times 10^{-5}}{F} \pi .
\]
The short-range risk per minute at one meter $\S$  is the volume of saliva inhaled times the probability $\pi$
that the source is infected. So we get
\[
\S  \ = \ 3.7  \times 10^{-6} \pi .
\]
When we divide these the $\pi$'s cancel, and we get the estimate for $\rho$,
the ratio of the long-range risk from aerosol transmission to short-range risk from facing someone
at 1 meter to be
\begin{equation}
\label{eq2}
\rho \ = \ \L/\S \ \approx \ \frac{6}{F} .
\end{equation} 

Sneezing releases much greater quantities of droplets of all sizes. 
Using the same methodology and Duguid's count of the number of droplets of different sizes produced by a sneeze, we get that 30 times the volume will be aerosolized from a sneeze than from counting from 1 to 100.
If the sneeze occurs once per minute, this would yield an increase by a factor of 50
\begin{equation}
\label{eqsn}
\rho_{\rm sneezing} \ \approx \ \frac{300}{F} .
\end{equation}
Of course, if an infected person is sneezing, the short-range risk also increases dramatically;
but (\ref{eqsn}) shows that even if the sneezing person is isolated at a sufficient distance that no sneezed
droplets hit anybody else, they still add greatly to the aerosol risk. (So, for example, if someone is alone in an elevator, sneezes, and then somebody else enters, they are at greatly increased risk).

Let us note that the recent paper \cite{Stadnytskyi11875} measures droplets emitted by speech, and comes up with a very much larger estimate than Duguid. They estimate that the total volume of emitted droplets
under 21 $\mu$m in diameter is 0.1 to 0.7 $\mu$L/minute, compared to the $10^{-4}$ $\mu$L/minute
estimate of Duguid. As the paper \cite{chen20} uses Duguid's results as input, if the actual number of droplets in
each size range scaled uniformly with Duguid's measurements, this would not affect the ratio in our conclusion. If Duguid undercounted small droplets more (which is almost certain), this would affect our conclusion.
Since most of the volume of the aerosolized droplets in  \cite{Stadnytskyi11875} comes from 
droplets in the 12-21 $\mu$m diameter range, and not from the large number but small volume of droplets
 in the below 4 $\mu$m range that were largely missed in older studies \cite{mor09}, we may hope that
 the correction does not completely invalidate our conclusion.
 (As \cite{Stadnytskyi11875} does not measure numbers of larger droplets, we cannot feed their
 data into the \cite{chen20} model).

To get an upper bound on $c$, we use the following calculation from \cite{chen20}.
They estimate that of all the droplets of diameter less than 50 $\mu$m emitted, the fraction
that enter the air-stream and are inhaled at 1m is $4.9 \times 10^{-3}$. Only counting these droplets, and not larger ones, understimates the short-range risk.
Let $R$ be the actual volume of aerosolized droplets emitted per minute that are under 50 $\mu$m in diameter (so $R$ is some fraction of $b$).
Let $q$ be the fraction of these droplets (calculated by volume) that become aerosolized.
Then from the estimates
\begin{eqnarray*}
\S &\ > \ &  4.9 \times 10^{-3}  R \pi \\
\L  & = & \frac{IRq\pi}{F}
\end{eqnarray*}
we get
\begin{equation}
\label{equb}
\rho \ < \ 2 \times 10^{2} \frac{Iq}{F} .
\end{equation}
Taking $I = 10$ as before, and noting that we must have $q \leq 1$, we get
\begin{equation}
\label{eqc}
c \ < \ 2000 .
\end{equation}

We get estimates, then, that $c$ is between $6$ and $2000$.
Since $q$ is likely to be much smaller than $1$ (unless nearly all of the volume of exhaled droplets
is in the sub 10 micron range), we expect that $c$ will turn out to be much closer to 6 than to 2000.

\section{Other comparisons}

To get a feel for what this says in practice, let us take $c = 10$ and $I$ = 10 in this section.

 A recommended air exchange rate is 8 liters per person per second, which is 480 liters per person per minute.
  If this were applied, 
 we would get $\rho = 0.02$ at full occupancy, and lower at reduced occupancy rates.

 

In \cite{Lu20} a restaurant in Guangzhou is studied, where on January 24, 2020, 5 people
sitting at adjacent tables to a person who was infected developed CoVid-19. 
This case has also been described and analyzed  in  \cite[Sec. 33]{EMG} and \cite{Buon20}. The air conditioning  did not
vent the air, so the air-exchange rate was very low (estimated at 0.56-0.77 exchanges per hour).
The relevant part of the restaurant had 21 people, and an area of 45 $m^2$. If we assume the volume is
90 $m^3$, this gives an $F$ of about 48 L/min.
This would yield a $\rho$ of $.2$ if the infected person were not coughing or sneezing.
Since this occurred before the pandemic was recognized, it is possible that the infected person did cough or
sneeze during the meal, which would greatly increase $\rho$ as noted above.

Elevators have different ventilation requirements. Some states require ``adequate means of ventilation'', without specifying what adequate is.
The state of Washington \cite{wasmv} requires them to have a ventilation of 1 cubic foot/minute per square foot (which is 1 foot/minute).
 At 1.5 square feet per person in the cab, 1 cfm/${\rm ft}^2$ would yield an $F$ of 42 L/min, and a $\rho$ of $.24$.

\section{Limitations}
Our result (\ref{eqlrv}) assumes perfect mixing. More likely is that being closer to the sources elevates the risk. It also assumes that the equilibrium level is reached quickly. This will over-estimate the risk especially in 
high-ceilinged environments for short durations.

The lack of direct measurements of inhaled saliva at different distances from a source,
coupled with uncertainty on how much virus load becomes aerosolized, required us to make 
 a string of assumptions to try to find $c$ in (\ref{eqma}). We do not believe that the value of $c$
as 6 is close to correct---we wanted to suggest one way to get at the order of magnitude,
which as explained above we would guess to be of the order of $10^1$.
 We would welcome
any improvements to this estimate.

\section{Notation}

$A$ 	\hs Total number of virions in aerosols in the air

$b$   \hs	Hydrated volume of droplets exhaled by an individual talking for one minute.

$E$	\hs Total volume of clean air brought in by the ventilation system (in Liters per minute)

$F$  \hs	Total volume of clean air brought in by the ventilation system per person in attendance (in Liters per minute) $ =  \frac{E}{n_t}$

$I$	\hs Breathing rate (in Liters per minute)

$\L$ \hs Long range risk = Number of aerosolized virions inhaled per minute, assuming infectious people are talking, not coughing or sneezing

$\L_{\rm sneezing}$ \hs Number of aerosolized virions inhaled per minute, assuming infectious people are sneezing once per minute

$k$ \hs	Fraction of the air that is replaced by the ventilation system per minute

$n_i$	\hs Number of infectious people present

$n_t$ \hs 	Total number of people present

$q$\hs  Fraction of emitted droplets under 50 $\mu$m in diameter that become aerosolized (weighted by volume)

$R$ \hs Hydrated volume of droplets smaller than 50 $\mu$m exhaled by an individual talking for one minute.

$\S$ \hs Short range risk = Number of virions inhaled per minute when talking face to face with an infectious person

$t$	 \hs Time (measured in minutes)

$v$ \hs	Number of aerosolized virions produced by one infectious  person per minute

$V$   \hs Total volume of enclosure (in Liters)

$\pi$ \hs Fraction of attendants who are infectious $ = \frac{n_i}{n_t}$

$\rho$ \hs The ratio $\L/\S$

$\rho_{\rm sneezing}$ \hs The ratio  $\L_{\rm sneezing} / \S$

\section{Disclosure of Interest Statement}

J.M., M.M. and B.D. received funding from Delaware North, a company that may be affected by the
research reported in the paper.

\bibliography{../Medical_references}

  \end{document}